# Density of States of an Electron in the Image Field and Blocking Electric Field

P.A. Golovinski, M.A. Preobrazhenskii, I.S. Surovtzev

Physics Research Laboratory, Voronezh State Technical University, 394 006 Voronezh, Russia
e-mail: golovinski@bk.ru

The motion of an electron in an image field and a blocking electric field is considered in semiclassical approximation. An exact analytical expression is found for the density of the energy spectrum of states. The dependence of spectral density on energy is obtained in a wide range of electric field strengths. The energy ranges with a qualitatively different structure of the spectrum are determined.

Near surface of a conductor, the charge is electrostatically attracted to its image. In the resulting one-dimensional potential, which is close in characteristics to the Coulomb field, bound states with Rydberg energy spectrum are formed: $\varepsilon_n = -1/32n^2$ a.u., $n = 1, 2, ..., \infty$ [1]. They manifest themselves experimentally in the resonances of the tunneling current [2] and photoemission spectra [3-6]. Bound near-surface states can significantly modify the response of a nanosystem to external field, therefore, the development of nanostructure researches stimulates the study of the influence of such states on various surface effects.

Earlier, in the framework of perturbation theory, the states of an electron bound by an image potential in external electric and magnetic fields were investigated [7]. The obtained analytical expressions for the probabilities of bound-bound and bound-free induced transitions made it possible to calculate the optical characteristics of an electron in the image potential [8]. Modern methods for generating terahertz unipolar pulses provide the creation of quasistatic fields with a strength of more than 1 MV/cm [9, 10]. In such strong fields, methods for describing the behavior of an electron based on perturbation theory are clearly insufficient. To calculate the states of an electron in an image field and a uniform electric field over a wide range of strength, we apply the semiclassical approach developed to describe the Stark effect in a hydrogen atom [11]. Using this method, wave functions and the spectrum of quasistationary states of an electron in the image potential was obtained when a constant destructive electric field was applied [12].

Of particular interest is the behavior of an electron bounded by an image field and a blocking electric field, which has no analogue in the three-dimensional problem. The resulting mechanism of one-dimensional electron confinement determines a completely discrete spectrum of states. The spatial width of the classically accessible region of electron motion is determined by its energy and field strength. The semiclassical calculations of the electron energy spectrum carried out earlier under such conditions were performed only for low field strength [13] in the region of negative energies. The aim of this work is to obtain the density of the spectrum of states of an electron in a wide range of field strengths and electron energy, including the vicinity of its zero value.

The stationary Schrödinger equation for an electron in an image field and a blocking uniform electric field directed perpendicular to the metal surface has the form

$$\left( \frac{\hat{p}_x^2 + \hat{p}_y^2 + \hat{p}_z^2}{2} - \frac{1}{4z} + Fz - E \right) \psi(x, x, y) = 0. \quad (1)$$

Here, the axis $z$ is directed perpendicular to the surface, the term $Fz$ in the Hamiltonian describes the interaction energy of an electron with uniform electric field acting on the electron with force, the $-1/4z$ term is the potential energy of the interaction of the electron with the image field of an ideal conductor. The variables in equation (1) are separated, so that a stationary solution we can search in the form $\psi = \varphi(z)\varphi_1(x)\varphi_2(y)$. The components $\varphi_1(x)$, $\varphi_2(y)$ of the wave function coincide with the eigenfunctions of the momentum operators, $\hat{p}_x, \hat{p}_y$, and are described by plane waves $\exp(ip_x x)$, $\exp(ip_y y)$. The component $\varphi(z)$ of the wave function is a solution to the equation

$$\frac{d^2\varphi(z)}{dz^2} + 2\left( \varepsilon - Fz + \frac{1}{4z} \right)\varphi(z) = 0, \quad (2)$$

and the total electron energy is

$$E = p_x^2/2 + p_y^2/2 + \varepsilon. \quad (3)$$

Figure 1 shows the dependence of the potential energy $U(z) = Fz - 1/4z$ of an electron on the distance to the surface of the metal at a field strength $F = 0.05$ a.u..

We describe the energy spectrum of equation (2) in the semiclassical approximation [13]. In potential $U(z)$, one-dimensional motion at any value $\varepsilon$ is finite, and the function

$$p(z) = \sqrt{2(\varepsilon - Fz + 1/4z)} \quad (4)$$

in the classically accessible region of motion $0 \le z \le z_1$ is coincided with the electron momentum component normal to the surface.

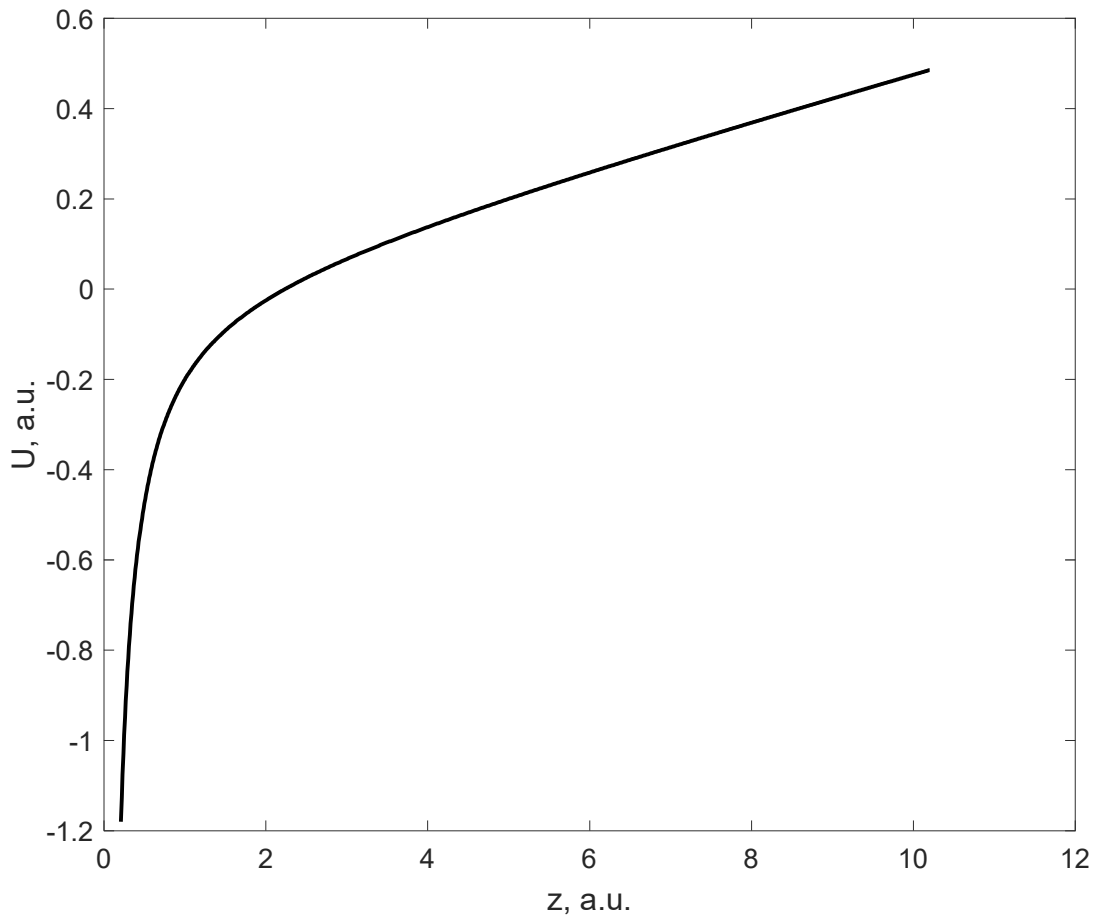

**Fig. 1.** The dependence of potential energy on the coordinate $z$.

The position of the classical turning point $z_1$ is determined by the positive root of the equation

$$z^2 - \varepsilon z / F - 1/4F = 0. \qquad (5)$$

Решения уравнения (5) следующим образом выражается через энергию электрона и напряженность поля:

$$z_{1,2} = \frac{\varepsilon}{2F} \pm \sqrt{\frac{\varepsilon^2}{4F^2} + \frac{1}{4F}}. \qquad (6)$$

Classic mechanical action

$$S = \int_0^{z_1} p(z)\,dz, \qquad (7)$$

allows us to write the quantization condition in the form [14]

$$S = \pi(n+1). \qquad (8)$$

Then, by differentiating relation (8) with respect to the quantum number gives an expression for the density of states.

$$\rho(\varepsilon) = \frac{dn}{d\varepsilon} = \frac{1}{\pi} \int_0^{z_1} \frac{dz}{p(z)}. \qquad (9)$$

When the external field is turned off, in the one-dimensional Coulomb potential a discrete spectrum is observed only at negative energy values. In this case, the density of states of the discrete spectrum for $\varepsilon < 0$ is determined by the expression

$$\frac{dn}{d\varepsilon} = \frac{|\varepsilon|^{-3/2}}{8\sqrt{2}}. \qquad (10)$$

This density of states has a singularity at an energy corresponding to the boundary of the continuous spectrum. When a blocking electric field is turned on, the electron motion becomes finite at any energy, and the spectrum is converted into a completely discrete one that does not have an infinite condensation point. In the limit of high electron energies, the influence of the Coulomb field can be neglected, and the model of the asymmetric triangular potential [15] becomes valid, in which the density of states can be calculated in the semiclassical approximation as

$$\rho(\varepsilon) = \frac{\sqrt{\varepsilon}}{2\sqrt{2}\,\pi F}. \qquad (11)$$

We now find the density of the spectrum while taking into account the image field and the blocking electric field. Momentum $p(z)$ is expressed through the roots of equation (5) in the form

$$p(z) = \sqrt{2F}\,\frac{\sqrt{(z-z_1)(z_2-z)}}{\sqrt{z}}, \qquad (12)$$

The integral in formula (9) in accordance with expression (12) has an integrable singularity at a point $z = z_1$, and the density of states is expressed in terms of complete elliptic integrals of the first and second kind [16]

$$K(q) = I_{-1/2}(q),\ E(q) = I_{1/2}(q),\ I_\mu(q) = \int_0^{\pi/2} (1 - q^2 \sin^2 x)^\mu\, dx \qquad (13)$$

in the form

$$\rho(\varepsilon) = \frac{2}{\pi\sqrt{2F(z_1 - z_2)}}\big((z_1 - z_2)E(q) + z_2 K(q)\big). \qquad (14)$$

Here the argument of elliptic integrals has the form $q = (z_1/(z_1 - z_2))^{1/2}$.

Figure 2 presents the results of a numerical calculation of the density of states at different values of the field strength. A sharp increase in the density of states is observed in the region of low modulo negative energies in the some interval, the width of which decreases with decreasing the field strength according to the estimation $\Delta\varepsilon \sim \sqrt{F}$. At positive energies $\varepsilon > \Delta\varepsilon$, the dependence rapidly attains of the asymptotic behavior of the triangular potential spectrum. The electron energy, at which the asymptotic behavior of the spectrum is realized and the density of the energy spectrum, decrease with increasing electric field strength.

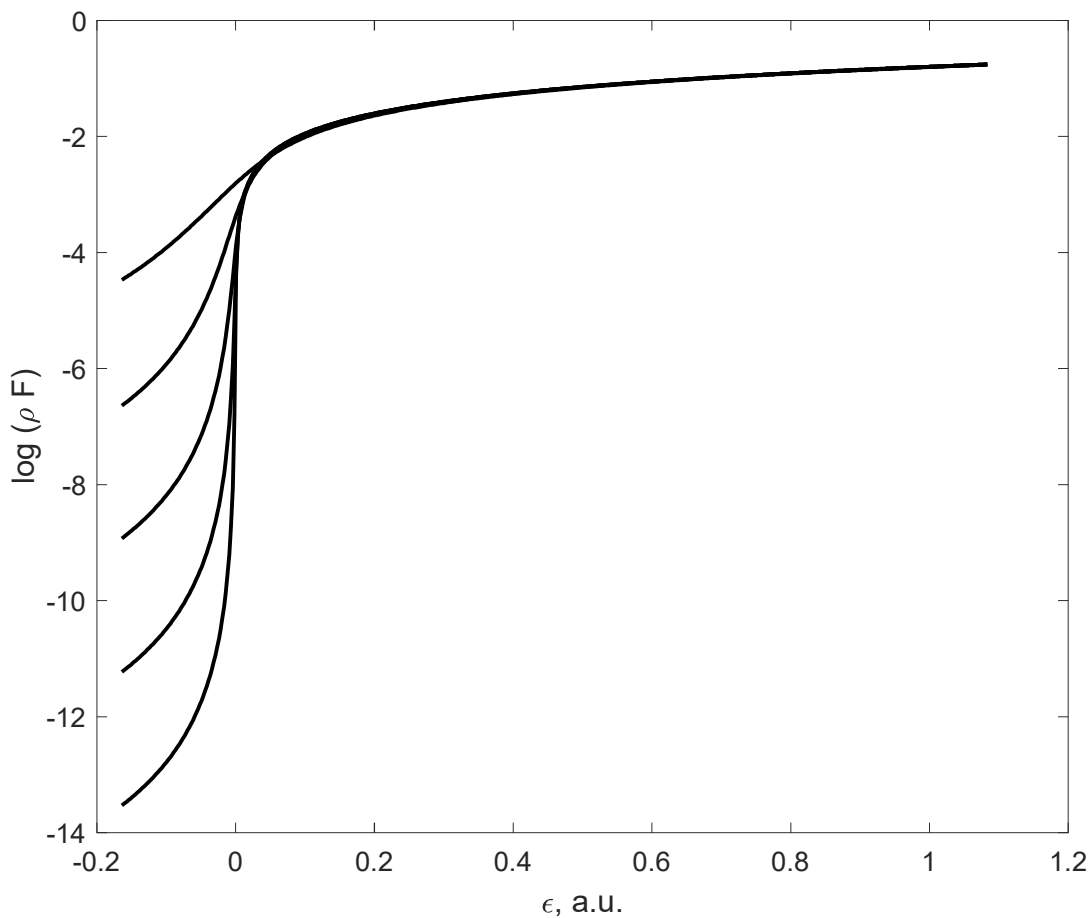

**Fig. 2.** The density of states multiplied by the value of the blocking field for fields with the strength $F = 10^{-2}, 10^{-3}, 10^{-4}, 10^{-5}, 10^{-6}$ a.u. depending on the energy. The curves are arranged from top to bottom in descending order of field strength.

Further studies will provide investigation the dynamics of an electron in an image field and constant electric field when exposed to ultrashort laser pulse.